\documentclass[natbib]{svjour3}
\usepackage{graphicx}
\usepackage{dcolumn}
\usepackage{bm}
\usepackage{latexsym}
\usepackage{amsmath,amssymb}
\usepackage[draft=false]{hyperref}
\usepackage[latin1]{inputenc}
\usepackage{latexsym}
\usepackage{amsmath}
\usepackage{ifpdf}
\usepackage{color}

\bibpunct{[}{]}{,}{n}{}{,} 

\RequirePackage{fix-cm}

\begin{document}

\title{Wormhole Cosmic Censorship
\thanks{We would like to thank Dar\'io N\'u\~nez for many helpful and useful
discussions and the "Laboratorio de Super-C\'omputo Astrof\'isico (LaSumA) del Cinvestav". 
This work was partially supported by PROMEP, DAIP-UGTO, and CONACyT M\'exico under grants CB-2011 no. 166212,
167335, and I0101/131/07 C-234/07 of the Instituto Avanzado de
Cosmologia (IAC) collaboration (http://www.iac.edu.mx/).
LAU-L was partially supported by PRODEP, DAIP, PIFI, CONACyT M\'exico under
grants 232893 (sabbatical), 167335 and 179881, Fundaci\'on Marcos
Moshinsky, and the Instituto Avanzado de Cosmolog\'ia (IAC)
collaboration}
}
\author{Tonatiuh Matos \and
L. Arturo Ure\~na-L\'opez \and
Galaxia Miranda}
\institute{T. Matos \and G. Miranda$^1$ \at 
			Departamento de F\'isica, Centro de Investigaci\'on y de
 			 Estudios Avanzados del IPN, A.P. 14-740, 07000 M\'exico D.F., M\'exico.\\
 			 \email{tmatos@fis.cinvestav.mx} \\
 			 \email{$^1$mmiranda@fis.cinvestav.mx}
			\and
			L. A. Ure\~na-L\'opez \at 			 
 			Departamento de F\'isica, DCI, Campus Le\'on, Universidad
 		 de Guanajuato, C.P. 37150, Le\'on, Guanajuato, M\'exico. \\
 		 \email{lurena@ugto.mx} 		 
 		 }

\date{Received: date / Accepted: date}

\maketitle

\begin{abstract}
We analyze the properties of a Kerr-like wormhole supported by phantom
matter, which is an exact solution of the Einstein-phantom field
equations. It is shown that the solution has a naked ring singularity
which is unreachable to null geodesics falling freely from
the outside. Similarly to Roger Penrose's cosmic censorship, that
states that all naked singularities in the Universe must be protected
by event horizons, here we conjecture from our results that a naked
singularity can also be fully protected by the intrinsic properties of
a wormhole's throat.
\end{abstract}

\keywords{Wormhole \and Singularity \and Cosmic Censorship}
\maketitle

Singularities are an ubiquitous ingredient in solutions of Einstein's
field equations \cite{Stephani:2003tm}. Most of the exact solutions known up to
date that represent local objects contain real singularities in their
spacetime structure. However, a singularity could be causally
connected with the rest of the Universe, and it is thought that any
number of troubles may come out of it. Besides, the predictability of
our physical theories breaks down since Einstein's equations are not
valid at the singularity. This is the reason why Roger Penrose
famously conjectured that spacetime singularities should be protected
by an event horizon that should in turn prevent external observers
from seeing them \cite{Penrose:1969pc,Penrose:1999vj}.

On the other hand, wormholes are one of the most interesting solutions
of Einstein
equations \cite{PhysRev.48.73,Morris:1988cz,Visser:1995cc}. They can be
seen as fast highways that connect separate places in the Universe, or
even as bridges between different universes. Unfortunately, such
solutions could exist provided that the matter source in Einstein's
equations can violate the null energy condition \cite{Morris:1988cz}.

More recently, phantom matter has emerged as a promising candidate to
be the dark energy in the Universe \cite{Clemson:2008ua}. At the local
level, phantom matter may be the matter source of wormholes
too, and then phantom wormholes are becoming again one of the most
mysterious and interesting solutions of Einstein's
equations \cite{Lobo:2005us}, see
also \cite{Gonzalez:2008wd,Gonzalez:2008xk,Gonzalez:2008zzd,Gonzalez:2009hn}.

In this Letter, we study in some detail the physical properties of the
Kerr-like wormhole found in Ref.\cite{Matos:2009au} and \cite{Miranda:2013gqa}, which is in fact
an exact solution of the Einstein's equations sourced by a phantom
scalar field. As we shall show, the relevant feature of the solution
is that its internal ring singularity is completely unreachable
because of the protection provided by the (special) wormhole's
throat. This is promising evidence that not only event horizons can
prevent external observers from seeing a singularity, but also can a
wormhole's throat.

To begin with, we write the line element describing the spacetime around the
wormholes in Boyer-Linquist coordinates as \footnote{We are using units
$c=1$. Notice also that metric~(\ref{eq:solBL}) is obtained from the
original one in Ref.\cite{Matos:2009au} and \cite{Miranda:2013gqa} through the change of
variables $l-l_1 \to l$.} \cite{Miranda:2013gqa}
\begin{equation}
  ds^2 = -f dt^2 + \frac{K}{f} dl^2 + \frac{\Delta_1}{f} \left[ K d\theta^2 +
    \sin^2\theta \, d\varphi^2 \right] \, ,  
\label{eq:solBL}
\end{equation}
where
\begin{subequations}
\begin{eqnarray}
   \label{eq:Kf}
  K = \frac{\Delta}{\Delta_1} \, , \quad f = \exp \left( - \frac{k_1}{2
      \Delta} \cos\theta \right)=\exp(-\lambda) \, , \\
  \label{eq:Deltas}
  \Delta = l^2 + l_0^2 \cos^2\theta \, ,  \quad \Delta_1 = l^2 + l_0^2
  \, , 
\end{eqnarray}
\end{subequations}
being $l_0$ a parameter with units of distance, and $k_1 > 0$ is a
parameter with units of angular momentum. The meaning of function
$f$ can be seen from the fact that the line element~(\ref{eq:solBL})
is an exact solution of the Einstein's equations, $R_{\mu\nu} = -8\pi
\, G \, \Phi_{\mu} \Phi_{\nu}$, where
\begin{equation}
  \Phi = \frac{1}{\sqrt{16 \pi G}} \, \lambda \, , 
  \label{eq:phantom}
\end{equation}
is a phantom-type scalar field \cite{Matos:2009au}. Moreover, the
Newtonian gravitational potential associated to
metric~(\ref{eq:solBL}) is given by $\phi_g = (1/2) \ln f$, and is
also directly related to the phantom field in Eq.~(\ref{eq:phantom}).

Some properties of metric~(\ref{eq:solBL}) are discussed in
turn. Firstly, for large values of the radial coordinate, $|l|
\gg l_0$, we have $\lambda \rightarrow 0$, $f \rightarrow 1$
and $\Delta, \Delta_1 \rightarrow l^2$, thus, the line
element~(\ref{eq:solBL}) is asymptotically flat at large distances,
\begin{equation}
ds^2 \rightarrow -dt^2 + dl^2 + l^2 \left( d\theta^2 +
\sin^2\theta\,d\varphi^2 \right) \, .
\label{eq:metric}
\end{equation}
However, as it usually happens with other wormholes, we must keep in
mind that there are two asymptotically flat regions, for $l \to \pm
\infty$, and the connection between them is called the throat of the
wormhole, which must be then located at $l=0$. Secondly, we can see
that $\Delta_1 > 0$ everywhere, but that that is not the case for
$\Delta$, which can be zero at different points of the
spacetime. Because of this, metric~(\ref{eq:solBL}) has a singularity
determined by the condition $\Delta=0$, which in turns translates into
a ring singularity described by $l = 0$ and $\theta = \pi/2$.

In order to verify that we have encountered a true singularity, we
should take a look at the invariants of the metric. For our case,
straightforward calculations show that the invariants can be generally
written as
\begin{equation}
  \text{Invariants} = \frac{F}{8 k^2_1 \Delta_1^{\alpha_2}
     k_1^{2\alpha_3}}\frac{ f^{2\alpha_3}}{\Delta^{\alpha_1} } \, ,
     \label{eq:invariants}
\end{equation}
where $F$ is a complicated function free of singularities that takes
different forms for each invariant of the metric, and $\alpha_1$,
$\alpha_2$, and $\alpha_3$ are positive coefficients whose exact value
depend upon the chosen invariant. It is nonetheless clear that the
condition $ f^{2\alpha_3}/\Delta^{\alpha_1}=\infty$ makes all of them 
diverge at the location of the ring singularity. 

\begin{figure}[htp!]
\includegraphics[width=8cm]{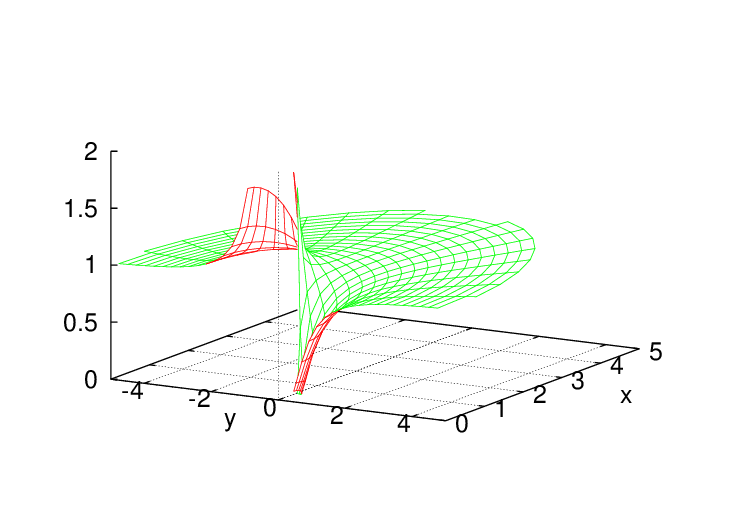}
\includegraphics[width=8cm]{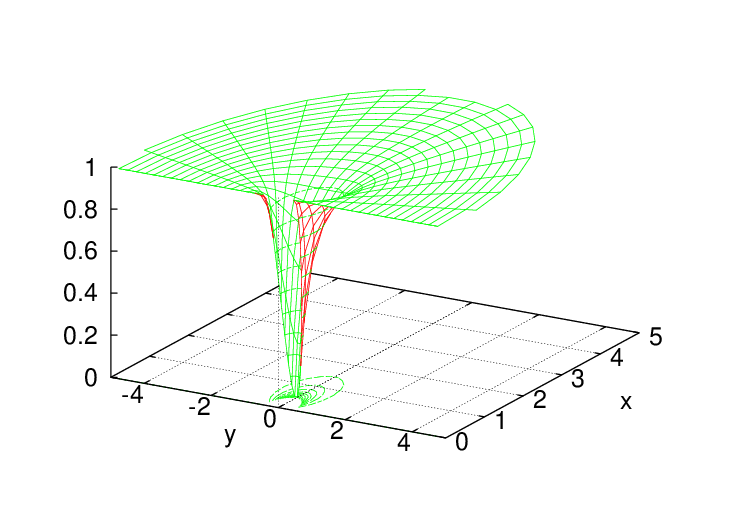}
\caption{The metric functions $f$ and $K$, see Eqs.~(\ref{eq:1}), for
  the values $l_{0}=0.45$ and $k_1=1$; the plot is given in terms of the
  pseudo-cartesian coordinates $(x,y)$ defined as $x= r \sin
  \theta$ and $y= r \cos \theta$, where the radial coordinate $r$ is
  defined as in Eq.~(\ref{eq:coordinates}). (Top) Function $f$ is the
  most troublesome metric function, as it shows quite different
  behaviors at the ring singularity, see also
  Eqs.~(\ref{eq:limits}). (Bottom) Function $K$ is regular everywhere,
  and it only vanishes at the ring singularity. At large distances,
  $|l| \gg l_0$, we recover the usual flat results for both functions,
  $f=1$ and $K=1$.
}
\label{fig:Kf}
\end{figure}

Two other metric quantities are affected by the ring
singularity. The most troublesome of them is function $f$, see
Eq.~(\ref{eq:Kf}), which is also shown in
Fig.~\ref{fig:Kf}. Notice that $f$ is discontinuous at the ring
singularity, as can be seen from the following limits:
\begin{subequations}
  \label{eq:limits}
  \begin{eqnarray}
    \lim_{\theta = \pi/2, |l| \to \infty} f = 1 \, , \quad
    \lim_{\theta = \pi/2, l \to 0} f = 1 \, , \\
    \lim_{l = 0, \theta \to (\pi/2)^+} f = \infty \, , \quad \lim_{l =
      0, \theta \to (\pi/2)^-} f = 0 \, .
  \end{eqnarray}
\end{subequations}

Observe that the only one function in metric~\eqref{eq:solBL} that
could be singular is $f$, all other elements are regular in the whole
of the space-time. Eqs.~\eqref{eq:limits} also tell us that the metric
function $f$ is singular if we reach the sphere $l=0$ through the
equatorial hypersurface $\theta=\pi/2$, and then this deserves a more
careful study. At $l=0$, the function
\begin{equation}
f_{\Delta}=\frac{ f^{2\alpha_3/\alpha_1}}{\Delta } =\exp\left(-\frac{k_1\alpha_3}{l_0^2\alpha_1\cos\theta}\right)\frac{1}{l_0^2\cos^2\theta}
  \end{equation}
has the limits
\begin{subequations}
  \label{eq:limits1}
  \begin{eqnarray}
    \lim_{\theta \to (\pi/2)^-}  f_{\Delta} &=& 0 \, , \quad 
    \lim_{\theta \to (\pi/2)^+} f_{\Delta} = \infty \,  .
  \end{eqnarray}
\end{subequations}
This also means that a singularity appears in the
invariants~(\ref{eq:invariants}) only if we reach the equator for
$\theta>\pi/2$. Thus, one observer can see the singularity only coming from
the southern hemisphere, as from the northern hemisphere the spacetime
is always regular. In conclusion, the singularity can only be reached
from the southern hemisphere $\theta \rightarrow \frac{\pi}{2}^+$.

On the other hand, the quantity inside the squared brackets in
Eq.~(\ref{eq:solBL}) can be interpreted as a solid angle element
modified by the metric function $K$:
\begin{equation}
  \label{metric2}
  d\Omega_0^2 = K d\theta^2 + \sin^2\theta \, d\varphi^2 \, ,
\end{equation}
and such a modified solid angle indicates that the spacetime is
indeed axially symmetric rather than spherically symmetric. Function
$K$ is well behaved everywhere, and it only vanishes at the location
of the ring singularity, $K(l=0, \theta = \pi/2)$ =0; at large
distances, $|l| \to \infty$, it takes on the usual flat result $K=1$
see Fig.~\ref{fig:Kf}.

To have a better visualization of the wormhole structure in
metric~(\ref{eq:solBL}), we define a new radial variable:
\begin{equation}
\label{eq:coordinates}
r^2 = \Delta_1 =  l^2 + l^2_0 \, ,
\end{equation}
and then write Eq.~(\ref{eq:solBL}) as a conformal metric:
\begin{equation}
  \label{eq:conformal}
    ds^2 = \frac{K}{f} \left( -\frac{f^2}{K} dt^2 + \frac{dr^2}{1
        -l^2_0/r^2} + \frac{r^2}{K} d\Omega^2_0 \right) \, ,
\end{equation}
where now we have
\begin{equation}
  \label{eq:1}
  K = 1 - \frac{l^2_0}{r^2} \sin^2 \theta \, , \quad f =
  \exp \left( - \frac{k_1}{2} \frac{\cos\theta}{r^2 - l^2_0 \sin^2
      \theta} \right) \, .
\end{equation}
It must be noticed that, in terms of the new variable, the ring
singularity is located at $\theta = \pi/2$ and $r=l_0$.

As we said before, the spacetime of the wormhole is axially symmetric,
and we can use this fact to draw up the structure of its throat by
taking a slice of the spacetime with $\varphi = 0$. In this case, the
induced metric, from Eq.~(\ref{eq:conformal}), is of the form $ds^2 =
(K/f) ds^2_c$, where
\begin{equation}
  \label{eq:morris-thorne}
  ds^2_c = -\frac{f^2}{K} dt^2 + \frac{dr^2}{1 - l^2_0/r^2} + r^2
  d\theta^2 \, .
\end{equation}
Curiously enough, $ds^2_c$ in Eq.~(\ref{eq:morris-thorne}) resembles the
line element of the famous Morris-Thorne (MT) wormhole, and then we can see
that there is a throat whose size is determined by the distance
parameter $l_0$ (see for
instance \cite{Morris:1988cz,Muller:2008zza}). This indicates that the
original metric represents a wormhole with a throat of finite size
that is conformally related to the MT solution.

There is, in addition, a nice property of conformal spaces that
states that the structure of null geodesics is preserved by
(non-singular) conformal transformations \cite{Wald:1984rg}. We can
then anticipate that null geodesics of metric~(\ref{eq:solBL}) are
conformally related to those of the MT wormhole (at least for the case
$\varphi=0$), through the conformal
metric~(\ref{eq:morris-thorne}). We just need to recall that null
geodesics of the MT wormhole, as drawn in terms of a typical embedding
diagram, follow paths that lie on the throat's surface, and this means
that our geodesics must likewise lie on such surface. In particular,
at distances far from the throat, we find that the conformal factor
$K/f \sim 1$, and then we shall recover exactly the case of the MT
wormhole.

However, there is the issue that the conformal transformation in our
case, see Eq.~(\ref{eq:conformal}), is not well behaved everywhere, as
the disturbing behavior of metric functions $K$, and $f$ shows up at
the ring singularity. Thus, the transformation should work well as
long as the problematic points, those of the ring singularity at
$r=l_0$ and $\theta = \pi/2$, are left out in our calculations.

We are to study now the null geodesics of test particles freely
falling into the wormhole. It proves convenient to work with the
Hamiltonian of the geodesics:
\begin{equation}
  2\mathcal{H} = - \frac{p_{t}^{2}}{f} +
  \frac{f\,p_{\varphi}^{2}}{\Delta_{1} \sin^2 \theta} + \frac{f}{K}
  \left( p_{l}^{2} + \frac{p_{\theta}^{2}}{\Delta _{1}} \right) \,
  , \label{eq:Hconstant}
\end{equation}
which is in itself a constant of motion, i.e. $\mathcal{H} = 0$ along
any given null geodesic.

Because the line element does not depend upon $t$ and $\varphi$
explicitly, their respective momenta are conserved, $p_t =
\mathrm{const.}$ and $p_\varphi = \mathrm{const.}$, whereas those
related to variables $l$ and $\theta$, $p_l$ and $p_\theta$,
respectively, will have a non-trivial dynamics. The Hamilton equations
of motion are
\begin{subequations}
  \label{eq:Hamiltonmotion}
  \begin{eqnarray}
    \label{eq:2}
    \dot{l} &=& \frac{f}{K} p_l \, , \quad \dot{\theta} =
    \frac{f}{\Delta} p_\theta \, , \\
    \dot{p}_l &=& - \frac{1}{2} \frac{\partial \ln f}{\partial l}
    \frac{p^2_t}{f} - \frac{1}{2} \frac{\partial \ln
      (f/\Delta_1)}{\partial l} \frac{f}{\Delta_1 \sin^2 \theta}
    p^2_\varphi \nonumber \\
    && - \frac{1}{2}  \frac{\partial \ln (f/K)}{\partial l}
    \frac{f}{K} p_{l}^{2} - \frac{1}{2} \frac{\partial \ln
      (f/\Delta)}{\partial l} \frac{f}{\Delta} p_{\theta}^{2} \, , \\
    \dot{p}_\theta &=& - \frac{1}{2} \frac{\partial \ln f}{\partial
      \theta} \frac{p^2_t}{f} - \frac{1}{2} \frac{\partial \ln
      (f/\sin^2 \theta)}{\partial \theta} \frac{f}{\Delta_1 \sin^2
      \theta}  p^2_\varphi \nonumber \\
    && - \frac{1}{2}  \frac{\partial \ln
      (f/K)}{\partial \theta} \frac{f}{K} \left( p_{l}^{2} +
      \frac{p_{\theta}^{2}}{\Delta _{1}} \right) \, ,
  \end{eqnarray}
\end{subequations}
where a dot denotes derivative with respect to an affine
parameter. Eqs.~(\ref{eq:Hamiltonmotion}) will be solved under
different initial conditions that will be set up at large distances
$l \gg l_0$. In every case the constraint $\mathcal{H} = 0$ will be
strictly accomplished, and as a general rule we will choose $p_t = 1$
and $p_l \leq 0$.

To have a connection with our previous discussion about the throat of
the wormhole, we find it useful to draw the paths of null geodesics
over the known embedding profile of the throat of a MT wormhole. In
order to achieve such a comparison, we set up $p_\varphi = 0$, and
will make use of the known embedding variable of the MT
wormhole \cite{Morris:1988cz}:
\begin{equation}
  \label{eq:transform}
  z(r) = l_0 \ln \left( \frac{r}{l_0} + \sqrt{\frac{r^2}{l^2_0} -1}
  \right) \, ,
\end{equation}
As we said before, in our case the geodesic paths will lie on the
throat's surface too, but will show deviations because of the
peculiarities induced upon them by the conformal factor $K/f$. Hence,
geodesic paths will give us a measure of the deformations in the true
throat of metric~(\ref{eq:solBL}) with respect to that of the MT's
wormhole.

\begin{figure}[htp!]
\includegraphics[width=9cm]{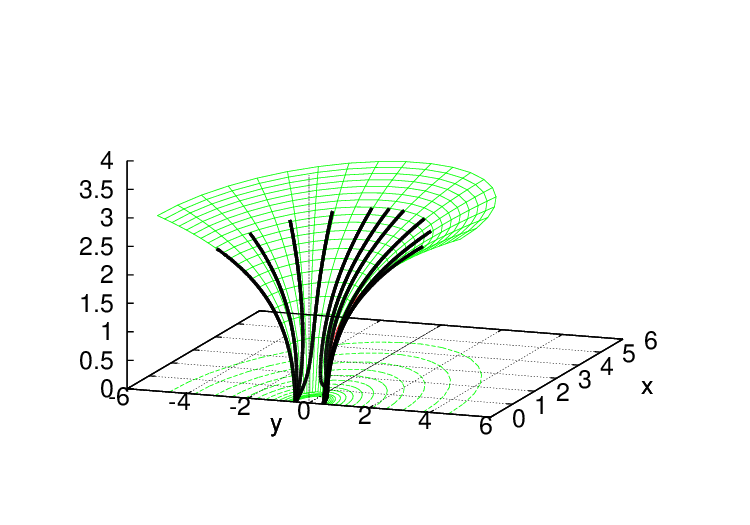}
\includegraphics[width=8cm]{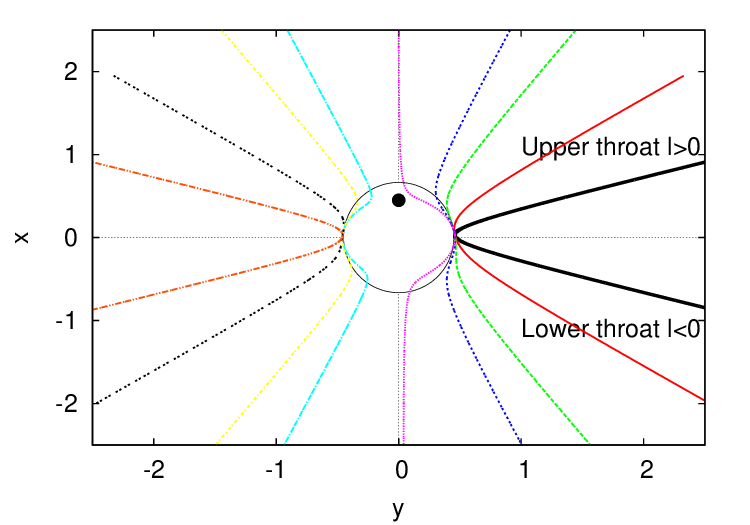}
\caption{Different null geodesics for different initial values of
  $\protect\theta$, given as numerical solutions of the
  geodesic equations~(\protect\ref{eq:Hamiltonmotion}); we are using
  here the pseudo-cartesian coordinates $x=r \sin \theta$ and $y=r \cos
  \theta$, whereas $z$ is given by the MT
  function~(\ref{eq:transform}). (Top) The surface represents the MT
  throat's profile, the solid lines are the null geodesics as depicted
  on the throat, and the ring singularity is marked as a point on the
  $xy$-plane at $r=l_0$ and $\theta = \pi/2$. (Bottom) The paths of
  null geodesics projected on the $xy$-plane; the upper half
  represents the upper part of the throat, $l > 0$ (also shown in
  the top plot), whereas the lower part represents the lower part of
  the throat, $l < 0$. In general, geodesics are able to avoid the
  naked singularity as they travel on the throat's surface. The
  deformation seen in the geodesics paths are due to the conformal
  factor $K/f$, see Eq.~(\ref{eq:morris-thorne}), that changes the
  real throat's surface with respect to MT's at points close to the
  ring singularity. See also the text for more details.}
\label{fig:geod1}
\end{figure}

The resulting trajectories are shown in Fig.~\ref{fig:geod1}. In
general terms, we can see that null geodesics are able to avoid the
naked singularity of the spacetime. This fact can be intuitively
understood from the conservation equation $\mathcal{H} =0$. For the
particular case considered in our numerical solutions shown in
Fig.~\ref{fig:geod1}, $p_t =1$ and $p_\varphi = 0$, and then
Eq.~(\ref{eq:Hconstant}) reads
\begin{equation}  
  \label{eq:barrier2}
  p^2_l + \frac{p^2_\theta}{r^2} = \frac{K}{f^2} \, .
\end{equation}
(Incidentally, this is the constraint equation that strictly
corresponds to the conformal metric~(\ref{eq:morris-thorne}).) As we
have seen before, at large distances $K/f^2 \sim 1$, and we recover
the usual equation of motion of the MT wormhole. However, as the
geodesics approach the singularity, we find that the term $K/f^2$
behaves discontinuously (a feature inherited from function $f$), and
then the constraint equation~(\ref{eq:barrier2}) cannot be satisfied
unless both momenta $p_l$ and $p_\theta$ are also discontinuous
there. There is no other option, for any finite solution, but to
deviate from the singularity point (see the bottom plot in
Fig.~\ref{fig:geod1}). In particular, the angular momentum $p_\theta$
is not conserved, and then the geodesic trajectory is scattered off
the singularity point: the closer the trajectory is to a singularity,
the larger the change in the angular momentum it acquires, and then
the larger the diversion the geodesic takes form the singularity.

At first sight, one could say that the naked singularity is protected
by an angular potential barrier. This is not completely true, because
the barrier is not build up from the conservation of angular momentum
(as is usually the case), but rather it is dynamically induced by the
same discontinuities of the metric functions. Moreover, these
discontinuities must be indeed related to the deformation of the
wormhole's throat in the neighborhood of the singularity, as indicated
by the geodesic paths drawn upon the MT wormhole's throat in
Fig.~\ref{fig:geod1}.

We would like to reinforce our geometrical point of view here. Even
though the conformal metric~(\ref{eq:morris-thorne}) resembles that of
the MT wormhole, there are some key differences. The first one is the
presence of a non-trivial gravitational potential: $\phi_g = (1/2) \ln
(f^2/K)$, which is the dynamical responsible of the deformation of
geodesic paths near the singularity. Hence, if
metric~(\ref{eq:morris-thorne}) were our only concern, then our
conclusion would be that the throat's surface is exactly that of the
MT wormhole, and that geodesic paths are deviated due to the presence
of the gravitational potential $\phi_g$.

However, the second key difference is that the conformal factor $K/f$ has
important effects on the form of the wormhole's throat too, and then
we can figure out that the throat is that of the MT wormhole plus
deformations in the neighborhood of the ring singularity. Therefore,
we must conclude that it is the form of the throat's surface which
prevents any null geodesic from reaching the ring singularity. In
other words, the reason behind the deformation of geodesic paths is
actually \emph{geometrical} rather than dynamical.

It is in this regard that our case resembles the case of an event
horizon surrounding the singularity at the centre of a black hole: it
is not a dynamical reason (i.e. the presence of a potential barrier)
that isolates the singularity, but a \emph{geometrical} event horizon
that changes the behavior of geodesic paths and separates out the
external and internal parts of the black hole.

In order to see the behavior of the geodesics close to the
singularity, we write the geodesic equations~(\ref{eq:Hamiltonmotion})
explicitly under the approximations $l<<l_0$ and $\theta\sim\pi/2$. We
obtain that:
  \begin{eqnarray}
      \dot p_l\frac{2\Delta}{f}
      &=&F_1\,l\,M-\frac{k_1l}{l_0^4\cos\theta}p_\varphi^2-\frac{k_1l}{l_0^2\cos\theta}\exp\left(\frac{k_1}{l_0^2\cos\theta}\right)p_t^2
      \, , \nonumber \\
    \dot p_{\theta} \frac{2\Delta}{f}
    &=&F_2\,M+\frac{k_1}{2l_0^2\sin\theta}p_\varphi^2+\frac{k_1}{2}\sin\theta\exp\left(\frac{k_1}{2l_0^2\cos\theta}\right)p_t^2
    \, , \nonumber \\
  \label{eq:HamiltonmotionLin}
  \end{eqnarray}
  being $M= p_l^2 l_0^2+ p_{\theta}^2$, and
  \[
  F_1=-\frac{ k_1-2\,l_0^2\cos \theta  }{l_0^4\cos^3 \theta}\, , \,\,\,\,\,F_2=\frac{  k_1-4l_0^2\cos \theta}{2 l_0^2 \cos^2\theta } \, .
  \]
Neglecting again the motion on the $\varphi$ direction, so that
$p_\varphi =0$, then Eqs.~(\ref{eq:HamiltonmotionLin}) can be
integrated exactly on the southern hemisphere as follows. We observe that $F_1l_0^2\sim
-F_{2,\theta}$, and then a solution of
Eqs.~\eqref{eq:HamiltonmotionLin} is
  \begin{eqnarray}
      p_l&=&A_0\sin(M_1ll_0)\exp\left(\frac{1}{2}F_1l^2\right)H 
      \, , \nonumber \\
    p_\theta&=&A_0\cos(M_1ll_0)\exp\left(\frac{1}{2}F_1l^2\right)H
    \, ,
    \label{eq:solution}
    \end{eqnarray}
where $M_{1,\theta}=F_1$, and
\begin{eqnarray*}
H=\exp\left(\frac{k_1}{4l_0^2\cos\theta}\right)\cos\theta \, .
\end{eqnarray*}

\begin{figure}[htp!]
\centering
\includegraphics[width=6cm]{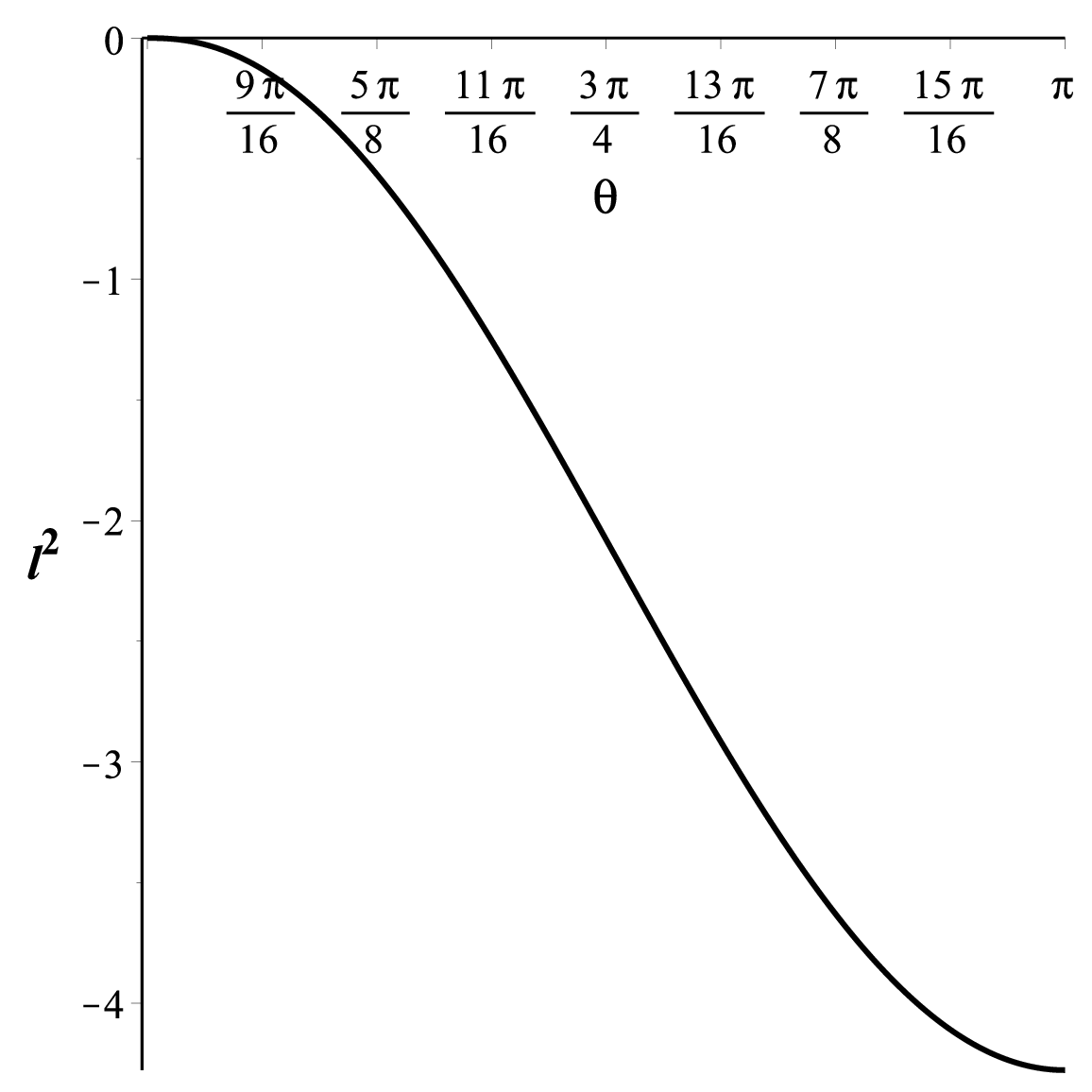}
\caption{Solution $l^2$ for the values $k_1=1$, $l_0=0.45$, $A_0=1$ and $p_t=5$.
In the region where the approximation is valid the function is negative. There are no solutions on the south hemisphere (near $\theta=\pi/2$) that are close to the ring singularity.}
\label{fig:l2}
\end{figure}

For metric~(\ref{eq:solBL}), any geodesic must comply with the
following constraint:
\begin{equation}
-\epsilon^2=-\frac{p_t^2}{f}+\frac{f}{\Delta}(p_l^2l_0^2+p_\theta^2)+\frac{f}{\Delta_1\sin^2\theta}
p_\varphi^2 \, , \label{eq:4}
\end{equation}
being $\epsilon^2=1$ for time-like geodesics, and $\epsilon^2=0$ for
null geodesics. Taking Eqs.~(\ref{eq:Kf}) and~(\ref{eq:solution})
into account, even for a null geodesic Eq.~\eqref{eq:4} transforms into
\begin{eqnarray}
l^2&=&-\frac{k_1l_0^2\cos^2\theta}{k_1-2l_0^2\cos\theta}
-\frac{l_0^4\cos^3\theta}{k_ 1-2l_0^2\cos\theta}\ln\left(\frac{p_t^2}{A_0^2}\right)\nonumber\\
 &\sim &-l_0^2\cos^2\theta\, . \label{eq:l2}
\end{eqnarray}
for $\theta\sim\pi/2^+$. It is clear that Eq.~\eqref{eq:l2} does not have
solutions for points on the southern hemisphere that are also close to
the ring singularity, as shown in Fig. \ref{fig:l2}. In other words, we see that it is not possible to
find a geodesic trajectory that can be in contact with the
singularity, and then the latter is not visible to exterior observers\cite{Deshingkar:2010ny}.

This leads us back again to the famous Penrose's cosmic
censorship conjecture. Despite admirable attempts to prove or disprove
it, its relevance in General Relativity is still a matter of debate,
as the conjecture has been proved to be true under special conditions
only \cite{Etesi:2012ag,Newman,Israel,Toth:2011ab,Wald1974548,Miyamoto:2011tt,Geroch:1979uc}. On the
other front line, there are counterexamples that violate the cosmic
censorship, like models of dynamical collapse leading to naked
singularities, in which trapped surfaces do not develop early enough
to shield them \cite{Joshi:2000ni,Bambi:2010mu}. Moreover, in \cite{Shapiro:1991zza} 
it was shown that naked singularities arise
in gravitational collapse, a result that has been debated by Penrose
himself \cite{Penrose:1999vj}. In any case, the possible violation of
the cosmic censorship has lead to new paths of study, like in the
analysis of the properties of naked singularities; for instance, its
observational consequences, and their differences from black holes
through lensing
data \cite{Batic:2010bz,Virbhadra:2007kw,Virbhadra:2002ju}.

The case of the wormhole we have studied throughout this Letter is
equally interesting for the discussions taking place in the
specialized literature. According to our solution, it is also possible
to protect a singularity if we surround it with a wormhole's
throat. This may indicate a generalization of Penrose's cosmic
censorship conjecture: it can be the case that naked singularities
with a more involved configuration may find the formation of a
wormhole's throat around them more convenient than just the appearance
of an event horizon.
%



%


\begin{thebibliography}{30}%
\makeatletter
\providecommand \@ifxundefined [1]{%
 \@ifx{#1\undefined}
}%
\providecommand \@ifnum [1]{%
 \ifnum #1\expandafter \@firstoftwo
 \else \expandafter \@secondoftwo
 \fi
}%
\providecommand \@ifx [1]{%
 \ifx #1\expandafter \@firstoftwo
 \else \expandafter \@secondoftwo
 \fi
}%
\providecommand \natexlab [1]{#1}%
\providecommand \enquote  [1]{``#1''}%
\providecommand \bibnamefont  [1]{#1}%
\providecommand \bibfnamefont [1]{#1}%
\providecommand \citenamefont [1]{#1}%
\providecommand \href@noop [0]{\@secondoftwo}%
\providecommand \href [0]{\begingroup \@sanitize@url \@href}%
\providecommand \@href[1]{\@@startlink{#1}\@@href}%
\providecommand \@@href[1]{\endgroup#1\@@endlink}%
\providecommand \@sanitize@url [0]{\catcode `\\12\catcode `\$12\catcode
  `\&12\catcode `\#12\catcode `\^12\catcode `\_12\catcode `\%12\relax}%
\providecommand \@@startlink[1]{}%
\providecommand \@@endlink[0]{}%
\providecommand \url  [0]{\begingroup\@sanitize@url \@url }%
\providecommand \@url [1]{\endgroup\@href {#1}{\urlprefix }}%
\providecommand \urlprefix  [0]{URL }%
\providecommand \Eprint [0]{\href }%
\providecommand \doibase [0]{http://dx.doi.org/}%
\providecommand \selectlanguage [0]{\@gobble}%
\providecommand \bibinfo  [0]{\@secondoftwo}%
\providecommand \bibfield  [0]{\@secondoftwo}%
\providecommand \translation [1]{[#1]}%
\providecommand \BibitemOpen [0]{}%
\providecommand \bibitemStop [0]{}%
\providecommand \bibitemNoStop [0]{.\EOS\space}%
\providecommand \EOS [0]{\spacefactor3000\relax}%
\providecommand \BibitemShut  [1]{\csname bibitem#1\endcsname}%
\let\auto@bib@innerbib\@empty
\bibitem [{\citenamefont {Stephani}\ \emph {et~al.}(2003)\citenamefont
  {Stephani}, \citenamefont {Kramer}, \citenamefont {MacCallum}, \citenamefont
  {Hoenselaers},\ and\ \citenamefont {Herlt}}]{Stephani:2003tm}%
  \BibitemOpen
  \bibfield  {author} {\bibinfo {author} {\bibfnamefont {H.}~\bibnamefont
  {Stephani}}, \bibinfo {author} {\bibfnamefont {D.}~\bibnamefont {Kramer}},
  \bibinfo {author} {\bibfnamefont {M.~A.}\ \bibnamefont {MacCallum}}, \bibinfo
  {author} {\bibfnamefont {C.}~\bibnamefont {Hoenselaers}}, \ and\ \bibinfo
  {author} {\bibfnamefont {E.}~\bibnamefont {Herlt}},\ }\href@noop {} {\emph
  {\bibinfo {title} {{Exact Solutions of Einstein's Field Equations.}}}}\
  \bibinfo {journal} {Second Edition.}
   \bibinfo {journal} {Cambridge University Press, Cambridge}
  (\bibinfo {year} {2003})\BibitemShut {NoStop}%
\bibitem [{\citenamefont {Penrose}(1969)}]{Penrose:1969pc}%
  \BibitemOpen
  \bibfield  {author} {\bibinfo {author} {\bibfnamefont {R.}~\bibnamefont
  {Penrose}},\ }\href@noop {} {\bibfield  {journal} {\bibinfo  {journal}
  {Riv.Nuovo Cim.}\ }\textbf {\bibinfo {volume} {1}},\ \bibinfo {pages} {252}
  (\bibinfo {year} {1969})}\BibitemShut {NoStop}%
\bibitem [{\citenamefont {Penrose}(1999)}]{Penrose:1999vj}%
  \BibitemOpen
  \bibfield  {author} {\bibinfo {author} {\bibfnamefont {R.}~\bibnamefont
  {Penrose}},\ }\href@noop {} {\bibfield  {journal} {\bibinfo  {journal}
  {J.Astrophys.Astron.}\ }\textbf {\bibinfo {volume} {20}},\ \bibinfo {pages}
  {233} (\bibinfo {year} {1999})}\BibitemShut {NoStop}%
\bibitem [{\citenamefont {Einstein}\ and\ \citenamefont
  {Rosen}(1935)}]{PhysRev.48.73}%
  \BibitemOpen
  \bibfield  {author} {\bibinfo {author} {\bibfnamefont {A.}~\bibnamefont
  {Einstein}}\ and\ \bibinfo {author} {\bibfnamefont {N.}~\bibnamefont
  {Rosen}},\ }\href {\doibase 10.1103/PhysRev.48.73} {\bibfield  {journal}
  {\bibinfo  {journal} {Phys. Rev.}\ }\textbf {\bibinfo {volume} {48}},\
  \bibinfo {pages} {73} (\bibinfo {year} {1935})}\BibitemShut {NoStop}%
\bibitem [{\citenamefont {Morris}\ and\ \citenamefont
  {Thorne}(1988)}]{Morris:1988cz}%
  \BibitemOpen
  \bibfield  {author} {\bibinfo {author} {\bibfnamefont {M.}~\bibnamefont
  {Morris}}\ and\ \bibinfo {author} {\bibfnamefont {K.}~\bibnamefont
  {Thorne}},\ }\href {\doibase 10.1119/1.15620} {\bibfield  {journal} {\bibinfo
   {journal} {Am.J.Phys.}\ }\textbf {\bibinfo {volume} {56}},\ \bibinfo {pages}
  {395} (\bibinfo {year} {1988})}\BibitemShut {NoStop}%
\bibitem [{\citenamefont {Visser}(1995)}]{Visser:1995cc}%
  \BibitemOpen
  \bibfield  {author} {\bibinfo {author} {\bibfnamefont {M.}~\bibnamefont
  {Visser}},\ }\href@noop {} {\emph {\bibinfo {title} {{Lorentzian wormholes:
  From Einstein to Hawking}}}}\ (\bibinfo {year} {1995})\ \bibinfo {note}
  {published in Woodbury, USA: AIP (1995) 412 p}\BibitemShut {NoStop}%
\bibitem [{\citenamefont {Clemson}\ and\ \citenamefont
  {Liddle}(2009)}]{Clemson:2008ua}%
  \BibitemOpen
  \bibfield  {author} {\bibinfo {author} {\bibfnamefont {T.~G.}\ \bibnamefont
  {Clemson}}\ and\ \bibinfo {author} {\bibfnamefont {A.~R.}\ \bibnamefont
  {Liddle}},\ }\href {\doibase 10.1111/j.1365-2966.2009.14641.x} {\bibfield
  {journal} {\bibinfo  {journal} {Mon.Not.Roy.Astron.Soc.}\ }\textbf {\bibinfo
  {volume} {395}},\ \bibinfo {pages} {1585} (\bibinfo {year} {2009})},\ \Eprint
  {http://arxiv.org/abs/0811.4676} {arXiv:0811.4676 [astro-ph]} \BibitemShut
  {NoStop}%
\bibitem [{\citenamefont {Lobo}(2005)}]{Lobo:2005us}%
  \BibitemOpen
  \bibfield  {author} {\bibinfo {author} {\bibfnamefont {F.~S. N.}\ \bibnamefont
  {Lobo}},\ }\href {\doibase 10.1103/PhysRevD.71.084011} {\bibfield  {journal}
  {\bibinfo  {journal} {Phys.Rev.}\ }\textbf {\bibinfo {volume} {D71}},\
  \bibinfo {pages} {084011} (\bibinfo {year} {2005})},\ \Eprint
  {http://arxiv.org/abs/gr-qc/0502099} {arXiv:gr-qc/0502099 [gr-qc]}
  \BibitemShut {NoStop}%
\bibitem [{\citenamefont {Gonzalez}\ \emph
  {et~al.}(2009{\natexlab{a}})\citenamefont {Gonzalez}, \citenamefont
  {Guzman},\ and\ \citenamefont {Sarbach}}]{Gonzalez:2008wd}%
  \BibitemOpen
  \bibfield  {author} {\bibinfo {author} {\bibfnamefont {J. A.}~\bibnamefont
  {Gonzalez}}, \bibinfo {author} {\bibfnamefont {F. S.}~\bibnamefont {Guzman}}, \
  and\ \bibinfo {author} {\bibfnamefont {O.}~\bibnamefont {Sarbach}},\ }\href
  {\doibase 10.1088/0264-9381/26/1/015010} {\bibfield  {journal} {\bibinfo
  {journal} {Class.Quant.Grav.}\ }\textbf {\bibinfo {volume} {26}},\ \bibinfo
  {pages} {015010} (\bibinfo {year} {2009}{\natexlab{a}})},\ \Eprint
  {http://arxiv.org/abs/0806.0608} {arXiv:0806.0608 [gr-qc]} \BibitemShut
  {NoStop}%
\bibitem [{\citenamefont {Gonzalez}\ \emph
  {et~al.}(2009{\natexlab{b}})\citenamefont {Gonzalez}, \citenamefont
  {Guzman},\ and\ \citenamefont {Sarbach}}]{Gonzalez:2008xk}%
  \BibitemOpen
  \bibfield  {author} {\bibinfo {author} {\bibfnamefont {J. A.}~\bibnamefont
  {Gonzalez}}, \bibinfo {author} {\bibfnamefont {F. S.}~\bibnamefont {Guzman}}, \
  and\ \bibinfo {author} {\bibfnamefont {O.}~\bibnamefont {Sarbach}},\ }\href
  {\doibase 10.1088/0264-9381/26/1/015011} {\bibfield  {journal} {\bibinfo
  {journal} {Class.Quant.Grav.}\ }\textbf {\bibinfo {volume} {26}},\ \bibinfo
  {pages} {015011} (\bibinfo {year} {2009}{\natexlab{b}})},\ \Eprint
  {http://arxiv.org/abs/0806.1370} {arXiv:0806.1370 [gr-qc]} \BibitemShut
  {NoStop}%
\bibitem [{\citenamefont {Gonzalez}\ \emph {et~al.}(2008)\citenamefont
  {Gonzalez}, \citenamefont {Guzman},\ and\ \citenamefont
  {Sarbach}}]{Gonzalez:2008zzd}%
  \BibitemOpen
  \bibfield  {author} {\bibinfo {author} {\bibfnamefont {J. A.}~\bibnamefont
  {Gonzalez}}, \bibinfo {author} {\bibfnamefont {F. S.}~\bibnamefont {Guzman}}, \
  and\ \bibinfo {author} {\bibfnamefont {O.}~\bibnamefont {Sarbach}},\ }\href
  {\doibase 10.1063/1.3058571} {\bibfield  {journal} {\bibinfo  {journal} {AIP
  Conf.Proc.}\ }\textbf {\bibinfo {volume} {1083}},\ \bibinfo {pages} {208}
  (\bibinfo {year} {2008})}\BibitemShut {NoStop}%
\bibitem [{\citenamefont {Gonzalez}\ \emph
  {et~al.}(2009{\natexlab{c}})\citenamefont {Gonzalez}, \citenamefont
  {Guzman},\ and\ \citenamefont {Sarbach}}]{Gonzalez:2009hn}%
  \BibitemOpen
  \bibfield  {author} {\bibinfo {author} {\bibfnamefont {J. A.}~\bibnamefont
  {Gonzalez}}, \bibinfo {author} {\bibfnamefont {F. S.}~\bibnamefont {Guzman}}, \
  and\ \bibinfo {author} {\bibfnamefont {O.}~\bibnamefont {Sarbach}},\ }\href
  {\doibase 10.1103/PhysRevD.80.024023} {\bibfield  {journal} {\bibinfo
  {journal} {Phys.Rev.}\ }\textbf {\bibinfo {volume} {D80}},\ \bibinfo {pages}
  {024023} (\bibinfo {year} {2009}{\natexlab{c}})},\ \Eprint
  {http://arxiv.org/abs/0906.0420} {arXiv:0906.0420 [gr-qc]} \BibitemShut
  {NoStop}%
\bibitem [{\citenamefont {Matos}(2010)}]{Matos:2009au}%
  \BibitemOpen
  \bibfield  {author} {\bibinfo {author} {\bibfnamefont {T.}~\bibnamefont
  {Matos}},\ }\href {\doibase 10.1007/s10714-010-0976-6} {\bibfield  {journal}
  {\bibinfo  {journal} {Gen.Rel.Grav.}\ }\textbf {\bibinfo {volume} {42}},\
  \bibinfo {pages} {1969} (\bibinfo {year} {2010})},\ \Eprint
  {http://arxiv.org/abs/0902.4439} {arXiv:0902.4439 [gr-qc]} \BibitemShut
  {NoStop}%
\bibitem [{\citenamefont {Miranda}\ \emph {et~al.}(2014)\citenamefont
  {Miranda}, \citenamefont {Matos},\ and\ \citenamefont
  {Montelongo-Garcia}}]{Miranda:2013gqa}%
  \BibitemOpen
  \bibfield  {author} {\bibinfo {author} {\bibfnamefont {G.}~\bibnamefont
  {Miranda}}, \bibinfo {author} {\bibfnamefont {T.}~\bibnamefont {Matos}}, \
  and\ \bibinfo {author} {\bibfnamefont {N.}~\bibnamefont
  {Montelongo-Garcia}},\ }\href@noop {} {\bibfield  {journal} {\bibinfo
  {journal} {Gen. Rev. Grav.}\ }\textbf {\bibinfo {volume} {46}},\ \bibinfo
  {pages} {1613} (\bibinfo {year} {2014})},\ \Eprint
  {http://arxiv.org/abs/1303.2410} {arXiv:1303.2410 [gr-qc]} \BibitemShut
  {NoStop}%
\bibitem [{\citenamefont {Muller}(2008)}]{Muller:2008zza}%
  \BibitemOpen
  \bibfield  {author} {\bibinfo {author} {\bibfnamefont {T.}~\bibnamefont
  {Muller}},\ }\href {\doibase 10.1103/PhysRevD.77.044043} {\bibfield
  {journal} {\bibinfo  {journal} {Phys.Rev.}\ }\textbf {\bibinfo {volume}
  {D77}},\ \bibinfo {pages} {044043} (\bibinfo {year} {2008})}\BibitemShut
  {NoStop}%
\bibitem [{\citenamefont {Wald}(1984)}]{Wald:1984rg}%
  \BibitemOpen
  \bibfield  {author} {\bibinfo {author} {\bibfnamefont {R.~M.}\ \bibnamefont
  {Wald}},\ }\href@noop {} {\emph {\bibinfo {title} {{General Relativity}}}}\
\bibinfo {journal} {Chicago, Usa: Univ. Pr.}
  (\bibinfo {year} {1984})\BibitemShut {NoStop}%
\bibitem [{\citenamefont {Deshingkar}(2010)}]{Deshingkar:2010ny}%
  \BibitemOpen
  \bibfield  {author} {\bibinfo {author} {\bibfnamefont {S.~S.}\ \bibnamefont
  {Deshingkar}},\ }\href@noop {} {\  (\bibinfo {year} {2010})},\ \Eprint
  {http://arxiv.org/abs/1012.3090} {arXiv:1012.3090 [gr-qc]} \BibitemShut
  {NoStop}%
\bibitem [{\citenamefont {Etesi}(2013)}]{Etesi:2012ag}%
  \BibitemOpen
  \bibfield  {author} {\bibinfo {author} {\bibfnamefont {G.}~\bibnamefont
  {Etesi}},\ }\href {\doibase 10.1007/s10773-012-1407-0} {\bibfield  {journal}
  {\bibinfo  {journal} {Int.J.Theor.Phys.}\ }\textbf {\bibinfo {volume} {52}},\
  \bibinfo {pages} {946} (\bibinfo {year} {2013})},\ \Eprint
  {http://arxiv.org/abs/1205.4550} {arXiv:1205.4550 [gr-qc]} \BibitemShut
  {NoStop}%
\bibitem [{\citenamefont {Newman}(1984)}]{Newman}%
  \BibitemOpen
  \bibfield  {author} {\bibinfo {author} {\bibfnamefont {R.}~\bibnamefont
  {Newman}},\ }\href {\doibase 10.1007/BF00762446} {\bibfield  {journal}
  {\bibinfo  {journal} {General Relativity and Gravitation}\ }\textbf {\bibinfo
  {volume} {16}},\ \bibinfo {pages} {175} (\bibinfo {year} {1984})}\BibitemShut
  {NoStop}%
\bibitem [{\citenamefont {Israel}(1984)}]{Israel}%
  \BibitemOpen
  \bibfield  {author} {\bibinfo {author} {\bibfnamefont {W.}~\bibnamefont
  {Israel}},\ }\href {\doibase 10.1007/BF01882488} {\bibfield  {journal}
  {\bibinfo  {journal} {Foundations of Physics}\ }\textbf {\bibinfo {volume}
  {14}},\ \bibinfo {pages} {1049} (\bibinfo {year} {1984})}\BibitemShut
  {NoStop}%
\bibitem [{\citenamefont {Toth}(2012)}]{Toth:2011ab}%
  \BibitemOpen
  \bibfield  {author} {\bibinfo {author} {\bibfnamefont {G.~Z.}\ \bibnamefont
  {Toth}},\ }\href {\doibase 10.1007/s10714-012-1374-z} {\bibfield  {journal}
  {\bibinfo  {journal} {Gen.Rel.Grav.}\ }\textbf {\bibinfo {volume} {44}},\
  \bibinfo {pages} {2019} (\bibinfo {year} {2012})},\ \Eprint
  {http://arxiv.org/abs/1112.2382} {arXiv:1112.2382 [gr-qc]} \BibitemShut
  {NoStop}%
\bibitem [{\citenamefont {Wald}(1974)}]{Wald1974548}%
  \BibitemOpen
  \bibfield  {author} {\bibinfo {author} {\bibfnamefont {R.}~\bibnamefont
  {Wald}},\ }\href {\doibase 10.1016/0003-4916(74)90125-0} {\bibfield
  {journal} {\bibinfo  {journal} {Annals of Physics}\ }\textbf {\bibinfo
  {volume} {82}},\ \bibinfo {pages} {548 } (\bibinfo {year}
  {1974})}\BibitemShut {NoStop}%
\bibitem [{\citenamefont {Miyamoto}\ \emph {et~al.}(2013)\citenamefont
  {Miyamoto}, \citenamefont {Jhingan},\ and\ \citenamefont
  {Harada}}]{Miyamoto:2011tt}%
  \BibitemOpen
  \bibfield  {author} {\bibinfo {author} {\bibfnamefont {U.}~\bibnamefont
  {Miyamoto}}, \bibinfo {author} {\bibfnamefont {S.}~\bibnamefont {Jhingan}}, \
  and\ \bibinfo {author} {\bibfnamefont {T.}~\bibnamefont {Harada}},\ }\href
  {\doibase 10.1093/ptep/ptt027} {\ ,\ \bibinfo {pages} {053E01} (\bibinfo
  {year} {2013})},\ \Eprint {http://arxiv.org/abs/1108.0248} {arXiv:1108.0248
  [gr-qc]} \BibitemShut {NoStop}%
\bibitem [{\citenamefont {Geroch}\ and\ \citenamefont
  {Horowitz}(1979)}]{Geroch:1979uc}%
  \BibitemOpen
  \bibfield  {author} {\bibinfo {author} {\bibfnamefont {R.~P.}\ \bibnamefont
  {Geroch}}\ and\ \bibinfo {author} {\bibfnamefont {G.}~\bibnamefont
  {Horowitz}},\ }\href@noop {} {\emph {\bibinfo {title} {{Global structure of
  spacetimes}}}}\ (\bibinfo {year} {1979})\ \bibinfo {note} {published in
  `General Relativity : An Einstein Centenary Survey'. Edited by S.W. Hawking,
  W. Israel. Cambridge University Press}\BibitemShut {NoStop}%
\bibitem [{\citenamefont {Joshi}(2000)}]{Joshi:2000ni}%
  \BibitemOpen
  \bibfield  {author} {\bibinfo {author} {\bibfnamefont {P.}~\bibnamefont
  {Joshi}},\ }\href@noop {} {\emph {\bibinfo {title} {{Gravitational Collapse:
  The Story so far}}}}\ (\bibinfo {year} {2000})\ \bibinfo {note} {published in
  `The Universe: Visions and Perspectives'. Edited by N. Dadhich and A.
  Kembhavi. Kluwer Academic Publishers. pp. 161-168}\BibitemShut {NoStop}%
\bibitem [{\citenamefont {Bambi}(2011)}]{Bambi:2010mu}%
  \BibitemOpen
  \bibfield  {author} {\bibinfo {author} {\bibfnamefont {C.}~\bibnamefont
  {Bambi}},\ }\href {\doibase 10.1088/1742-6596/283/1/012005} {\bibfield
  {journal} {\bibinfo  {journal} {J.Phys.Conf.Ser.}\ }\textbf {\bibinfo
  {volume} {283}},\ \bibinfo {pages} {012005} (\bibinfo {year} {2011})},\
  \Eprint {http://arxiv.org/abs/1008.3026} {arXiv:1008.3026 [gr-qc]}
  \BibitemShut {NoStop}%
\bibitem [{\citenamefont {Shapiro}\ and\ \citenamefont
  {Teukolsky}(1991)}]{Shapiro:1991zza}%
  \BibitemOpen
  \bibfield  {author} {\bibinfo {author} {\bibfnamefont {S.~L.}\ \bibnamefont
  {Shapiro}}\ and\ \bibinfo {author} {\bibfnamefont {S.~A.}\ \bibnamefont
  {Teukolsky}},\ }\href {\doibase 10.1103/PhysRevLett.66.994} {\bibfield
  {journal} {\bibinfo  {journal} {Phys.Rev.Lett.}\ }\textbf {\bibinfo {volume}
  {66}},\ \bibinfo {pages} {994} (\bibinfo {year} {1991})}\BibitemShut
  {NoStop}%
\bibitem [{\citenamefont {Batic}\ \emph {et~al.}(2011)\citenamefont {Batic},
  \citenamefont {Chin},\ and\ \citenamefont {Nowakowski}}]{Batic:2010bz}%
  \BibitemOpen
  \bibfield  {author} {\bibinfo {author} {\bibfnamefont {D.}~\bibnamefont
  {Batic}}, \bibinfo {author} {\bibfnamefont {D.}~\bibnamefont {Chin}}, \ and\
  \bibinfo {author} {\bibfnamefont {M.}~\bibnamefont {Nowakowski}},\ }\href
  {\doibase 10.1140/epjc/s10052-011-1624-3} {\bibfield  {journal} {\bibinfo
  {journal} {Eur.Phys.J.}\ }\textbf {\bibinfo {volume} {C71}},\ \bibinfo
  {pages} {1624} (\bibinfo {year} {2011})},\ \Eprint
  {http://arxiv.org/abs/1005.1561} {arXiv:1005.1561 [gr-qc]} \BibitemShut
  {NoStop}%
\bibitem [{\citenamefont {Virbhadra}\ and\ \citenamefont
  {Keeton}(2008)}]{Virbhadra:2007kw}%
  \BibitemOpen
  \bibfield  {author} {\bibinfo {author} {\bibfnamefont {K. S.}~\bibnamefont
  {Virbhadra}}\ and\ \bibinfo {author} {\bibfnamefont {C. R.}~\bibnamefont
  {Keeton}},\ }\href {\doibase 10.1103/PhysRevD.77.124014} {\bibfield
  {journal} {\bibinfo  {journal} {Phys.Rev.}\ }\textbf {\bibinfo {volume}
  {D77}},\ \bibinfo {pages} {124014} (\bibinfo {year} {2008})},\ \Eprint
  {http://arxiv.org/abs/0710.2333} {arXiv:0710.2333 [gr-qc]} \BibitemShut
  {NoStop}%
\bibitem [{\citenamefont {Virbhadra}\ and\ \citenamefont
  {Ellis}(2002)}]{Virbhadra:2002ju}%
  \BibitemOpen
  \bibfield  {author} {\bibinfo {author} {\bibfnamefont {K.~S.}\ \bibnamefont
  {Virbhadra}}\ and\ \bibinfo {author} {\bibfnamefont {G.~F.~R.}\ \bibnamefont
  {Ellis}},\ }\href {\doibase 10.1103/PhysRevD.65.103004} {\bibfield  {journal}
  {\bibinfo  {journal} {Phys.Rev.}\ }\textbf {\bibinfo {volume} {D65}},\
  \bibinfo {pages} {103004} (\bibinfo {year} {2002})}\BibitemShut {NoStop}%
\end{thebibliography}
\end{document}